\documentstyle[12pt,aaspp4,flushrt,epsf]{article}

\def\be{\begin{equation}}
\def\ee{\end{equation}}
\def\mathnew{\mathsurround=0pt}
\def\simov#1#2{\lower .5pt\vbox{\baselineskip0pt \lineskip-.5pt
        \ialign{$\mathnew#1\hfil##\hfil$\crcr#2\crcr\sim\crcr}}}

\def\half{{\textstyle{1\over2}}}
\def\bfJ{{\bf J}}
\def\bfw{{\bf w}}
\def\p{\partial}
\def\bfOmega{{\bf \Omega}}
\def\bfH{{\bf H}}
\def\bfA{{\bf A}}
\def\bfD{{\bf D}}
\def\bfy{{\bf y}}
\def\bfx{{\bf x}}

\def\bflambda{\mbox{\boldmath $\lambda$}}
\def\ffrac#1#2{{\textstyle\frac{#1}{#2}}}

\begin{document}

\title{The geometry of phase mixing}

\author{Scott Tremaine}
\affil{Princeton University Observatory, Peyton Hall, Princeton, NJ~08544 \\
        tremaine@astro.princeton.edu}

\begin{abstract}

Partially phase-mixed structures in galaxies occupy a complex surface of
dimension $D$ in six-dimensional phase space. The appearance of such
structures to observers is determined by their projection into a subspace
whose dimensionality $K$ is determined by the number of observables (e.g. sky
position, distance, radial velocity, etc.).  We discuss the expected
dimensionality of phase-space structures and suggest that the most prominent
features in surveys with $K\ge D$ will be stable singularities
(catastrophes). The simplest of these are the shells seen in the outer parts
of elliptical galaxies.

\end{abstract}

\section{Introduction}

\noindent
The evolution of the phase-space density of stars in galaxies is determined by
the collisionless Boltzmann equation, which states that phase-space flow
is incompressible.  Thus a cloud of stars in phase space
becomes more and more distorted as it evolves; the local or fine-grained
density around any point in phase space remains the same, but the
coarse-grained density evolves towards a stationary state. This process, known
as phase mixing, is the principal mechanism by which stellar systems reach
coarse-grained equilibrium.  The standard analogy is stirring a glass
containing 20\% rum and 80\% Coke; eventually every finite volume element in
the glass contains 20\% rum and 80\% Coke, even though infinitesimal volume
elements are either 100\% rum or 100\% Coke (e.g. \cite{AA68}).

Galaxies are not thoroughly phase-mixed, since they are at most a few
dynamical times old in their outer parts, and since mergers and star formation
continuously add new stars. There is a variety of direct observational
evidence for incomplete phase mixing: (i) Sharp-edged features (``shells'') in
the outer parts of at least 10\% of elliptical galaxies are believed to arise
from the recent tidal disruption of small galaxies (e.g. \cite{HQ88}). (ii)
Proper-motion and radial-velocity surveys of the local halo reveal clumping in
phase space (\cite{MHM96}). (iii) The tidally disrupted Sagittarius dwarf
galaxy (\cite{IWG97}) provides a concrete example of a cloud in the early
stages of phase mixing. (iv) Moving groups in the solar neighborhood may
result from dissolution and mixing of star clusters and associations
(\cite{E65,D98}).

Incomplete phase mixing is likely to play a growing role in the interpretation
of observations, for several reasons: radial-velocity and proper-motion
surveys are rapidly improving in size and quality; large telescopes with
improved image quality will permit us to examine the surface-brightness
structure of elliptical galaxies with high spatial resolution and
signal-to-noise ratio; the Hipparcos mission has dramatically improved the
precision of measurements of the phase-space distribution of stars in the
solar neighborhood, and future space-based astrometric missions such as SIM
and GAIA will provide proper-motion and parallax measurements over much of the
Galaxy; and the Sloan Digital Sky Survey will soon provide a stellar database
of unprecedented size and uniformity ($10^8$ stars with five-color
photometry).

\begin{figure}
\vspace{10cm}
\includegraphics{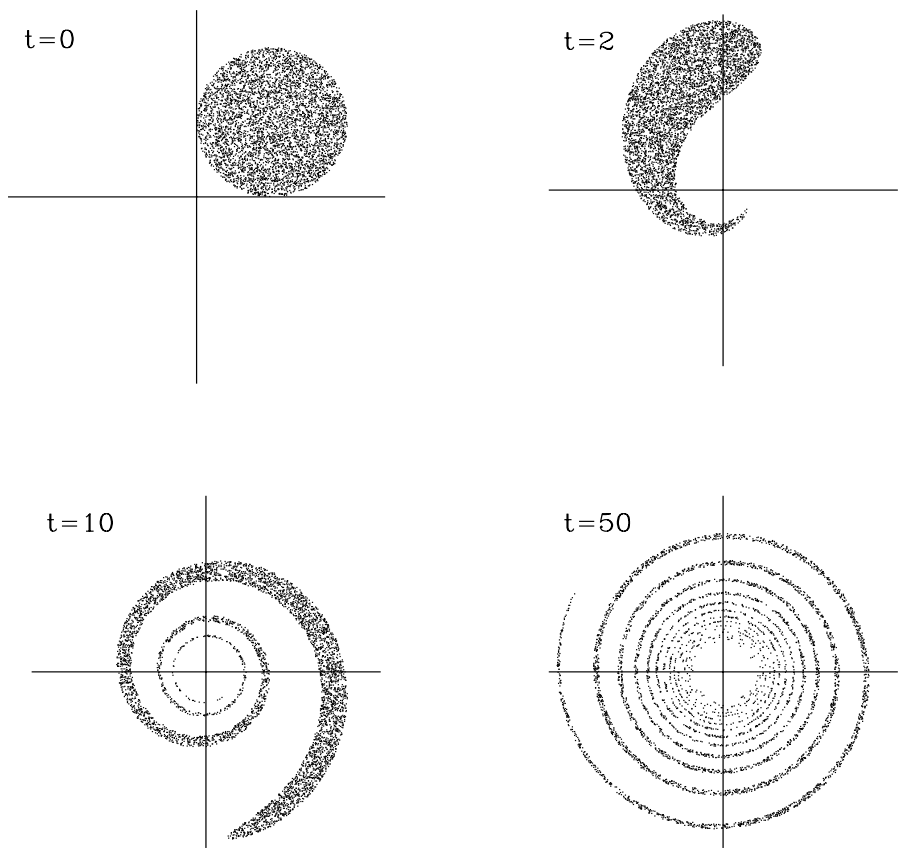}
\caption{Phase mixing in a two-dimensional phase space. The figure shows the
evolution of 5000 points following the equation of motion $\dot x=-y/r$, $\dot
y=x/r$ where $r^2=x^2+y^2$.}
\label{fig:mix}
\end{figure}

Virtually all theoretical descriptions of phase mixing have focused on mixing
in a two-dimensional phase space, which is the simplest to visualize
(Fig. \ref{fig:mix}).  This paper discusses some of the geometrical features
of partially phase-mixed systems (we shall call these ``phase structures'')
with higher dimensionality, and the projection of this geometry into the
observables that are measured in a survey. We shall argue that this geometry
can be organized by two integers: the dimension of the phase structure and the
dimension of the survey.

\section{The dimension of phase structures}

\label{sec:dim}

\noindent
A cloud of stars orbiting in an integrable potential (the ``host galaxy'') can
be viewed as a surface or manifold of dimension $D\le N$ embedded in
$N$-dimensional phase space (usually $N=6$). For example, a cold disk has
$D=2$ (the disk occupies a two-dimensional surface in configuration space, and
the velocity at each position is unique), collisionless cold dark matter has
$D=3$ (three spatial dimensions but zero random velocity), and a hot galaxy
has $D=6$. Since real stellar systems always have
non-zero thickness and velocity dispersion, we are referring to an
``effective'' or ``coarse-grained'' dimension; loosely speaking, a structure
in a phase space of $N$ dimensions has effective dimension $D$ if its extent
in $D$ independent directions is much larger than its extent in the other
$N-D$ directions.

Because the flow of the cloud through phase space is incompressible, the
dimension of the manifold does not change in the initial stages of phase
mixing, although the shape of the manifold becomes more and more
complicated. However, the effective dimension can change over longer times,
through several distinct processes. The most obvious is that the phase-mixing
scale grows smaller and smaller with time (see Figure \ref{fig:mix}), and
eventually becomes smaller than the coarse-graining scale. At this point the
phase structure has the same (coarse-grained) dimension as the phase space,
unless all of the stars in the cloud have the same value of one or more
isolating integrals of the motion. Gravitational scattering by small-scale
irregularities (e.g. massive objects) can also lead to diffusion of the phase
structure and an increase in effective dimension.

Less obviously, phase mixing can also {\it decrease} the number of effective
dimensions. For example, consider the cloud resulting from the disruption of a
small hot galaxy. The motion of the stars in the cloud can be described by a
Hamiltonian $H(\bfJ)$, where $(\bfw,\bfJ)$ are action-angle variables in the
host potential. The equations of motion are
\be
\bfJ=\hbox{const},\qquad {d\bfw\over dt}={\p H(\bfJ)\over\p \bfJ}\equiv
\bfOmega(\bfJ).
\ee
The trajectory of the centroid of the cloud is $(\bfJ_c(t),\bfw_c(t))=
(\bfJ_0,\bfw_0+\bfOmega_0t)$, where $\bfOmega_0\equiv\bfOmega(\bfJ_0)$ and
$(\bfJ_0,\bfw_0)$ is approximately the location of the center of the satellite
at the disruption time $t=0$.

Similarly, the trajectory of a star in the cloud is
$(\bfJ_0+\Delta\bfJ,\bfw_0+\Delta\bfw+\bfOmega(\bfJ_0+\Delta\bfJ)t)$, where
$|\Delta\bfJ|\ll \bfJ_0$ and $|\Delta\bfw|\ll 1$. This can be simplified to
\be
(\bfJ(t),\bfw(t))=(\bfJ_c,
\bfw_c(t))+(\Delta\bfJ,\Delta\bfw+\bfH\Delta\bfJ t)+{\rm
O}(\Delta\bfJ,\Delta\bfw)^2,
\label{eq:hess}
\ee
where 
\be
H_{ik}=\p^2 H(\bfJ_0)/\p J_i\p J_k
\label{eq:hhh}
\ee
 is the Hessian of the
Hamiltonian. At large times the extent of the cloud is dominated by the terms
$\propto t$ in equation (\ref{eq:hess}), so we may write approximately
\be
\bfJ(t)\simeq \bfJ_0+{\rm O}(\Delta\bfJ), \qquad 
\bfw(t)\simeq \bfw_c(t)+\bfH\Delta\bfJ t+{\rm O}(\Delta \bfw). 
\label{eq:hessa}
\ee 
Since $\bfH$ is symmetric, it is diagonalizable, that is, there exists an
orthogonal matrix $\bfA$ such that 
\be 
\bfA\bfH\bfA^{-1}=\bfD({\bflambda}),
\ee 
where $D_{ij}({\bflambda})=\lambda_i\delta_{ij}$, $\bflambda=\{\lambda_i\}$
are the eigenvalues of $\bfH$, and $\bfA^t=\bfA^{-1}$. We now make a canonical
transformation to new action-angle variables $(\bfJ',\bfw')$ using the mixed
generating function $S(\bfJ',\bfw)=\bfJ'\bfA\bfw$; thus
\be 
\bfw'={\p S\over\p \bfJ'}=\bfA\bfw, \qquad \bfJ={\p S\over\p
\bfw}=\bfA^{t}\bfJ', \qquad \bfw=\bfA^t\bfw', \qquad \bfJ'=\bfA\bfJ.
\ee
Equation (\ref{eq:hessa}) now simplifies to \be J_i'(t)\simeq
J_{0i}'+{\rm O}(\Delta J'), \qquad w'_i(t)\simeq w_{ci}'(t)+\lambda_i\Delta
J'_i t+{\rm O}(\Delta w').
\label{eq:hessb}
\ee

This result shows that the shape of the expanding cloud is largely determined
by the eigenvalues $\bflambda$ of the Hessian $\bfH$, which are invariant
under the orthogonal transformation $\bfA$. If two of the three eigenvalues 
are zero (say $\lambda_2=\lambda_3=0$) then the cloud expands only
along the $w_1'$ direction in phase space, yielding a one-dimensional tidal
streamer that grows linearly with time. If one of the eigenvalues is zero (say
$\lambda_3=0$) then the cloud expands in two dimensions in phase
space. However, if in addition $|\lambda_1\Delta J_1'| \gg |\lambda_2 \Delta
J_2'|$ or vice versa, then the expansion will effectively be
one-dimensional. Thus the dimension $D$ has been reduced from 6 to
1 or 2, because the cloud expands much faster in some angles than in
others, and not at all in action space.

We illustrate these remarks with some examples. The Hamiltonian for the
triaxial harmonic oscillator may be written $H(\bfJ)=\sum_{k=1}^3\omega_kJ_k$,
where $\omega_k$ is the frequency along axis $k$. The Hessian for this
Hamiltonian is zero, so the disrupted system does not expand at all: the
frequencies are independent of the actions and phase mixing does not occur.
The Hamiltonian for the Kepler potential is $H(\bfJ)=-\half(GM)^2/J_1^2$,
where $J_1=(GMa)^{1/2}$ and $a$ is the semi-major axis. The eigenvalues of the
Hessian are $(-3/a^2,0,0)$, so the expanded cloud is one-dimensional, even if
its original state was 6-dimensional.  Simulations of disruptions of small
galaxies in a spherical logarithmic potential yield one-dimensional clouds as
well (\cite{JHB96}), but the precise conditions determining the effective
dimension of the cloud resulting from tidal disruption of a small stellar
system have not yet been studied.

\section{Catastrophes}

\noindent
We now survey the phase space by measuring $K$ of the $N$ phase-space
coordinates of each star (we call this $K$-dimensional space the ``observable
space''\footnote{Other names include control space or external variables.}),
and ask how to detect the $D$-dimensional phase structure. For example, an
image or set of sky positions has $K=2$; a data cube (positions and radial
velocities) has $K=3$; a proper-motion survey has $K=4$, etc. A bolometric
laboratory detector of cold dark matter measures only their energy and
moreover is restricted to a single point in configuration space.

If $D<K$ the phase structure will appear as a distinct entity in the
observable space, which can be detected by standard techniques: plotting the
data on paper ($K=2$), visualization software ($K=3$), cluster-finding
algorithms, etc. More specialized and powerful techniques can be used if the
phase structure has known properties: for example, the disruption of distant
satellite galaxies in a spherical potential leads to phase structures which
are great circles on the sky (\cite{LB95,JHB96}). 

In this discussion we concentrate the more challenging case when $D=K$, so
that the phase structure covers most or all of the observable space.  In this
case, the most prominent features in the survey will be the singularities of
the phase structure, that is, the locations where the projection of the phase
structure into the observable space leads to a singularity.  For example, if
in Figure \ref{fig:mix} the observations yield only the horizontal coordinate,
then near-singularities arise in the plot at $t=50$ when the curve
representing the phase structure is vertical.

We can generalize this simple concept to singularities with higher dimension
using catastrophe theory (\cite{PS78,BU80,G81}). We begin by
setting up a coordinate system $(u_1,\ldots,u_D)$ on the phase structure that
contains the cloud. We complete the coordinate system of the phase space by
$(u_{D+1},\ldots,u_N)$, which are chosen so that $u_{D+1}=\cdots=u_N=0$ on
the phase structure. Thus the phase structure is specified by
\be
{\p \phi({\bf u})\over \p u_j}=0, \quad j=1,\ldots,N\qquad\hbox{where}\qquad 
 \phi({\bf u})=\sum_{j=D+1}^N u_j^2. 
\ee
The advantage of specifying the phase structure in terms of the generating
function $\phi$ is that $\phi$ is a single-valued function of the phase-space
coordinates.

Next let $(x_1,\ldots,x_D)$ be the coordinates of the observable space, and
complete the coordinate system of the phase space by $(y_1,\ldots,y_{N-D})$ (we
call this $(N-D)$-dimensional space the ``hidden space''\footnote{Other names
include state space or internal variables.}).  For given values of the
observable coordinates, the location or locations of the phase structure are
specified implicitly by the equations 
\be 
{\p \phi(\bfy;\bfx)\over \p y_j}=0, \quad j=1,\ldots,N-D.
\label{eq:sing}
\ee
Singularities in the observable space arise when there are displacements in
the hidden space that leave (\ref{eq:sing}) unaltered, that is, for which
\be
\sum_{k=1}^{N-D}{\p^2\phi(\bfy;\bfx)\over\p y_j\p y_k}dx_k=0,
\qquad j=1,\ldots,N-D.
\ee
These equations have a non-trivial solution if the 
determinant of the $(N-D)\times(N-D)$ Hessian matrix vanishes, that is, if
\be
\hbox{det}{\p^2\phi(\bfy;\bfx)\over \p y_j\p y_k}=0.
\label{eq:hessmat}
\ee

We can restrict our attention to structurally stable singularities (roughly
speaking, ones for which small changes in $\phi$ lead to small changes in the
locus of the singularity); the reason is that unstable singularities represent
a subset of measure zero and hence are atypical, at least for
$N\le 7$ (\cite{Z77}). 

Catastrophes are structurally stable singularities of gradient maps. They are
classified by their codimension, which is the dimension of the observable
space minus the dimension of the singularity. Thus codimension 1 corresponds
to a singular point in a survey with 1 observable, a singular line in a survey
with 2 observables, etc. The corank of the catastrophe is the minimum value of
$N-D$, the dimensionality of the hidden space, which is required for the
singularity to occur. A third classification is by the degree $k$ of the
singularity: the mean density in a neighborhood of the observable space of
radius $\epsilon$ around the singularity is $\propto\epsilon^{-k}$.

\subsection{Fold}

\noindent
The fold is the only catastrophe with codimension 1. Since the corank of the
fold is also 1, the simplest example occurs in a phase space with two
dimensions, one of which is observable. For example, the generating function
\be 
\phi(y;x)=\ffrac{1}{3}y^3+xy 
\ee 
implies that the phase structure is the
parabolic curve $x+y^2=0$, and this manifold has a fold catastrophe at
$x=0$. If the linear density on the curve is uniform, the observable density
is 
\be 
\rho(x)\propto \left(1-{1\over 4x}\right)^{1/2}, \qquad x<0, 
\ee 
and zero otherwise. The square-root divergence in density 
on one side of the singularity is characteristic of a fold, and implies that
the degree $k=\half$. 

Examples of fold catastrophes occur in the projection of a spherical shell
onto a plane, rainbows, sunlight sparkling on the sea, twinkling of starlight,
gravitational lensing, etc.

The simplest examples of fold catastrophes in phase mixing are shell
structures in elliptical galaxies, which are believed to arise from disrupted
companion galaxies. Shells are one-dimensional structures that appear in
two-dimensional observable space (the two coordinates on the sky plane) and
hence if they are singularities they must have codimension 1 and hence must be
folds. Hernquist and Quinn (1988) \nocite{HQ88} carried out simulations
of the evolution of disrupted companions in
spherical host galaxy potentials, and several of their conclusions can be
interpreted in terms of the results we have derived so far. They find that
shell formation requires the accretion of either a dynamically cold or a
spatially compact companion; these are precisely the conditions required so
that the cloud dimension $D\le 2$ (\S \ref{sec:dim}), which in turn is
required so that singularities are present in a two-dimensional observable
space. They distinguish ``spatial folding'' (projection from three-dimensional
configuration space to two-dimensional space) from ``phase-wrapping''
(projection from phase space into three-dimensional space) but this
distinction is not fundamental: both are projections from phase space
into observable space. Finally, they point out that some shells arise from
step functions in surface density rather than square-root singularities; the
former are not singularities and hence are not described by catastrophe
theory. 

\subsection{Cusp}

\noindent
The cusp is the only catastrophe with codimension $2$, and has corank 1. The
cusp is a singular point in a 2-dimensional observable space, a line in
3-dimensional observable space, etc. For example, consider a phase space with
$N=3$ dimensions, two of which are observable. The generating function 
\be
\phi(y;\bfx)=\ffrac{1}{4}y^4+\half x_2y^2+x_1y
\label{eq:gencusp}
\ee
implies that the phase structure is given by the surface $y^3+x_2y+x_1=0$,
which is singular (in the sense of eq. \ref{eq:hessmat}) along the lines
\be
x_1=\pm {2\over 3^{3/2}}\left(-x_2\right)^{3/2}.
\label{eq:tan}
\ee
Each of these lines is a fold catastrophe, and their junction is a cusp
catastrophe.

If the surface density of stars is uniform on the phase structure, the surface
density in the observable space $(x_1,x_2)$ is given parametrically by 
\be
\Sigma(x_1,x_2)\propto\left[1+{1+y^2\over (x_2+3y^2)^2}\right]^{1/2},
\qquad\hbox{where}\qquad x_1=-y^3-x_2y. 
\label{eq:csig}
\ee At a distance $\Delta x_2$ from the fold ($|\Delta x_2|\ll |x_2|\sim
|x_1|^{2/3}\ll 1$), the density $\Sigma\propto |x_1|^{-2/3}$ for $\Delta
x_2>0$ and $\Sigma\propto |x_1|^{-1/3}(-\Delta x_2)^{-1/2}$ for $\Delta x_2<0$
(the characteristic fold behavior). The degree of the cusp singularity is
$k=\frac{2}{3}$.

\begin{figure}
\vspace{10cm}
\includegraphics{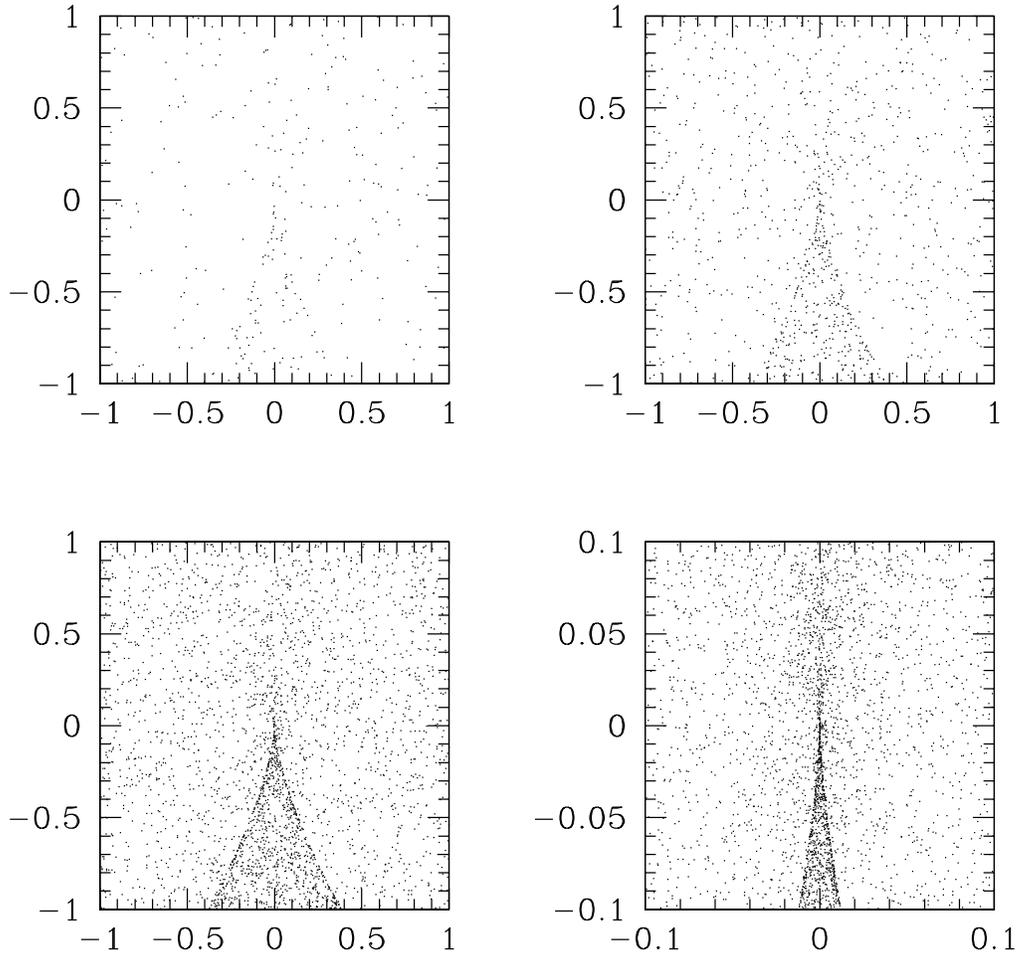}
\caption{The cusp catastrophe defined by the generating function
(\ref{eq:gencusp}), assuming a uniform density of stars on
the manifold defined by $\p\phi/\p y=0$. The first three panels show samples
of 300, 1000, and 3000 points over the square $[-1,1]^2$; the fourth panel
shows a sample of 3000 points over $[-0.1,0.1]^2$. The cusp point at the
origin unfolds into two fold catastrophes. }
\label{fig:one}
\end{figure}

To illustrate the appearance of a cusp catastrophe, the first three panels of
Figure \ref{fig:one} show a survey that samples the surface density
(\ref{eq:csig}) with 300, 1000, and 3000 data points in the interval
$x_1,x_2\in[-1,1]$. The fourth panel shows a higher-resolution survey with
3000 points in $x_1,x_2\in[-0.1,0.1]$.  The general cusp in a 2-dimensional
observable space is formed from this special case by smooth distortion
(i.e. diffeomorphism) of the observable space, and still consists of two fold
catastrophes meeting with a common tangent, as in equation (\ref{eq:tan}).

Examples of cusp catastrophes include the caustic curve seen on the surface of
coffee in a cup, and the critical point of the van der Waals equation of state
in the pressure-temperature plane.

One might speculate that some of the dwarf spheroidal satellite galaxies are
cusp catastrophes rather than bound equilibrium stellar systems, but their
isopleths are approximately elliptical (\cite{IH95}) and do not resemble the
density contours that are expected near a cusp.

\subsection{Catastrophes with codimension 3}

\noindent
There are three catastrophes with codimension $3$, the swallowtail, elliptic
umbilic, and hyperbolic umbilic. The swallowtail has corank 1, and thus first
appears in a phase space with $N\ge 4$ dimensions; the umbilics have corank 2
and thus require $N\ge 5$. All of these singularities are points in a
3-dimensional observable space, lines in 4-dimensional observable space,
etc. Just as a cusp occurs at the junction of two fold curves, these occur at
the junction of cusp curves, which are connected by fold surfaces.

\subsubsection{Swallowtail}

\noindent
Consider a phase space with $N=4$ dimensions, three of which are
observable. The generating function 
\be
\phi(y;\bfx)=\ffrac{1}{5}y^5+\ffrac{1}{3}x_3y^3+\half x_2y^2+
x_1y,
\label{eq:genswallow}
\ee
implies that the phase structure is given by $y^4+x_3y^2+x_2y+x_1=0$,
which is singular (in the sense of eq. \ref{eq:hessmat}) along the surfaces
defined in terms of the parameters $(y,x_3)$ by
\be
x_1=3y^4+x_3y^2,\qquad x_2=-2(2y^3+x_3y).
\label{eq:tans}
\ee
Each of these surfaces is a fold catastrophe; the folds meet at cusp lines defined
by
\be
x_1=-3y^4,\qquad x_2=8y^3, \qquad x_3=-6y^2.
\label{eq:cusps}
\ee
The fold surfaces intersect one another along the parabola $x_3^2=4x_1$.  The
degree of the swallowtail is $k=\frac{3}{4}$.

\begin{figure}
\vspace{8cm}
\includegraphics{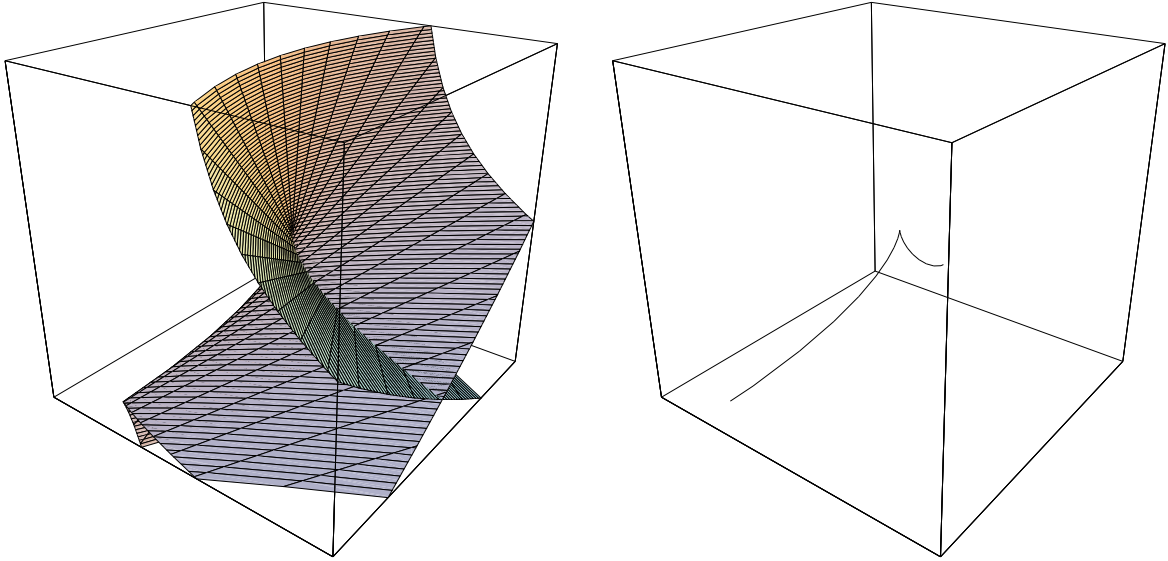}
\caption{The swallowtail catastrophe defined by the generating function
(\ref{eq:genswallow}). The left panel shows the fold surfaces and the right
panel shows the cusp lines.}
\label{fig:two}
\end{figure}

Figure \ref{fig:two} shows the fold surfaces and cusp lines for the
swallowtail catastrophe. 

\subsubsection{Elliptic umbilic}

\begin{figure}
\vspace{8cm}
\includegraphics{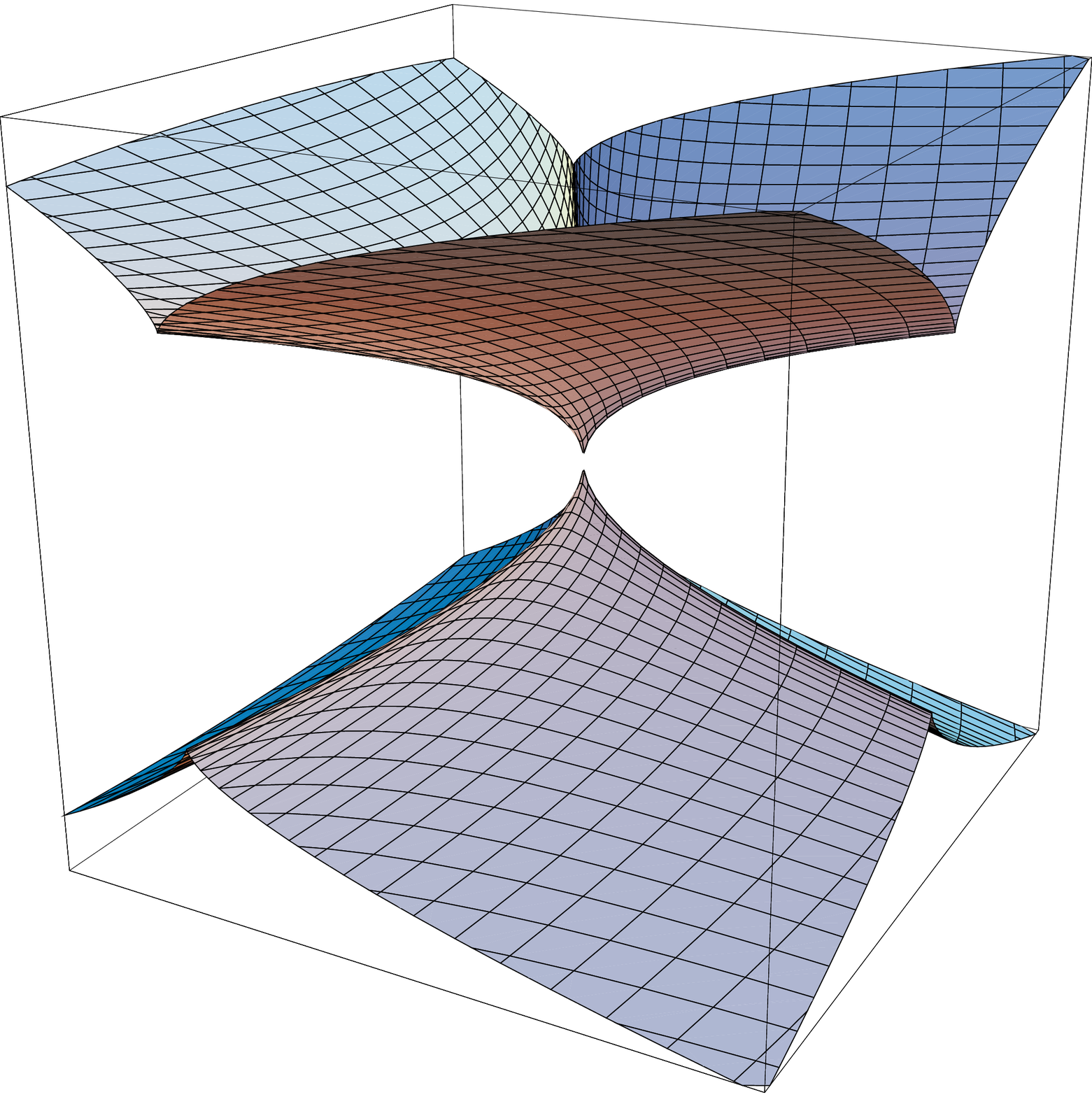}
\includegraphics{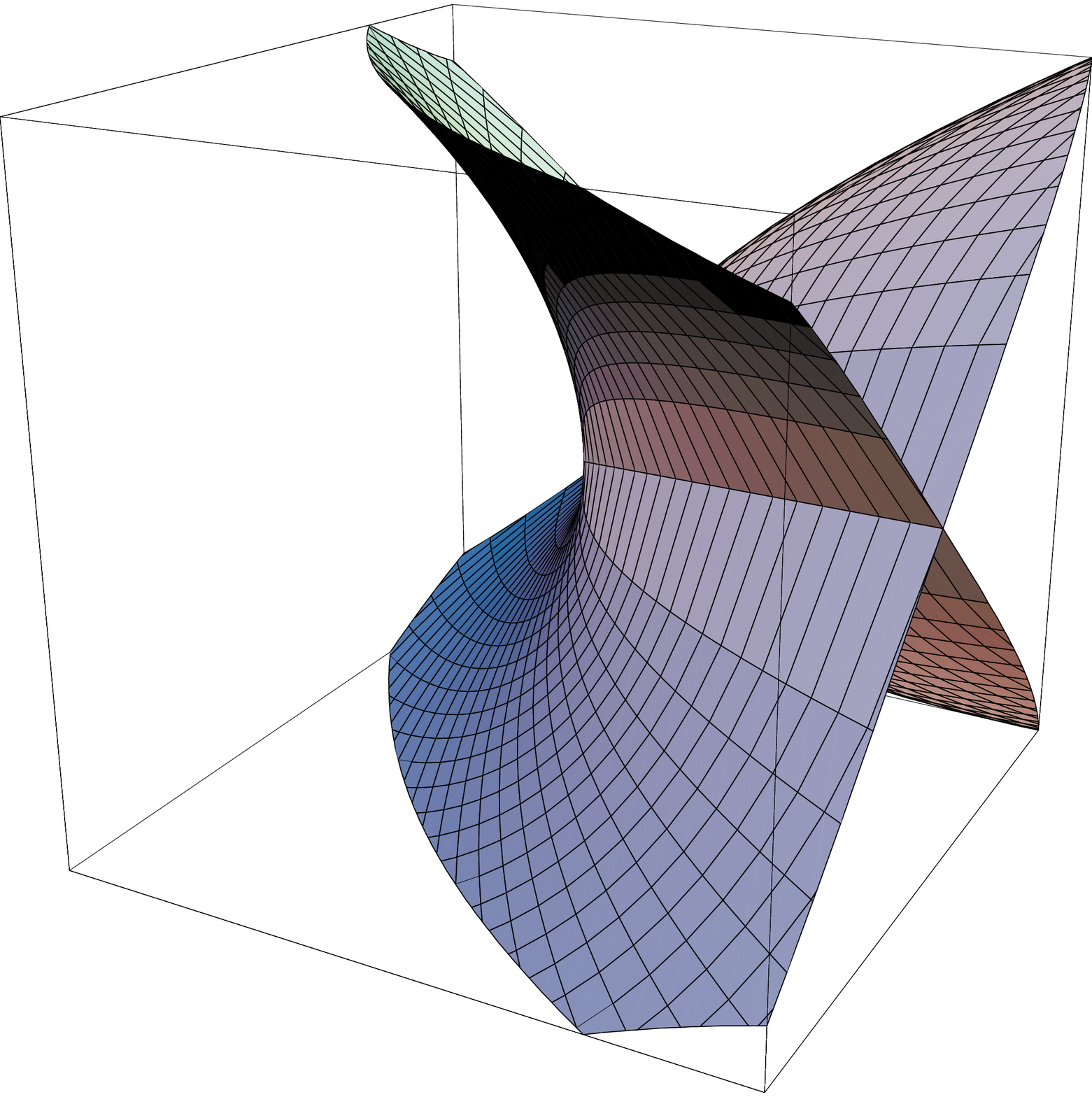}
\caption{The elliptic umbilic (left) and hyperbolic umbilic (right)
catastrophes.} 
\label{fig:three}
\end{figure}

\noindent
Consider a phase space with $N=5$ dimensions, three of which are
observable. The generating function \be
\phi(y_1,y_2;\bfx)=y_1^3-3y_1y_2^2-x_3(y_1^2+y_2^2)-x_2y_2-x_1y_1
\label{eq:genumbe}
\ee
implies that the phase structure is given by the surface
\be
3y_1^2-3y_2^2-2x_3y_1-x_1=0, \qquad 6y_1y_2+2x_3y_2+x_2=0. 
\ee
which is singular along the surfaces defined parametrically by
\be
x_1=3y_1^2-3y_2^2\pm 6y_1(y_1^2+y_2^2)^{1/2}, \quad
x_2=-6y_1y_2\pm 6y_2(y_1^2+y_2^2)^{1/2}, \quad
x_3=\mp(y_1^2+y_2^2)^{1/2}.
\ee
Each of these surfaces is a fold catastrophe; the folds meet at three
parabolic cusp lines defined by $x_1^2+x_2^2=x_3^4$ and confined to planes at
angles of $120\deg$ that intersect along the $x_3$-axis (see Figure
\ref{fig:three}). The degree is $k=1$.

\subsubsection{Hyperbolic umbilic}

\noindent
The generating function 
\be 
\phi(y_1,y_2;\bfx)=y_1^3+y_2^3-x_3y_1y_2-x_2y_2-x_1y_1
\label{eq:genumbh}
\ee
implies that the phase structure is given parametrically by
\be
x_1=3y_1^2+\epsilon_1(y_1y_2^3)^{1/2}, \quad
x_2=3y_2^2+\epsilon_1(y_1^3y_2)^{1/2}, \quad
x_3=\epsilon_2(y_1y_2)^{1/2},
\ee
where $\epsilon_i=\pm 1$ and $y_1,y_2\ge 0$. 
Each of these surfaces is a fold catastrophe; the folds intersect along
the coordinate axes $x_1>0,x_2=x_3=0$ and $x_1=0,x_2>0,x_3=0$
(Fig. \ref{fig:three}). The cusp line is defined parametrically by
\be
x_1=x_2=9y^2, \qquad x_3=6\epsilon_2y, \qquad y>0.
\ee
The degree is $k=1$.

These exhaust the catastrophes with codimension $\le 3$. Catastrophes with
higher codimension are only relevant when the observable space has dimension
$D\ge 4$. Catastrophes of higher codimension become increasingly complicated,
and are less important since we have argued in \S\ref{sec:dim} that typical
phase structures will often have dimension 3 or less. 

\section{Summary}

\noindent
It is important to understand the geometry of phase structures before
searching for them. We have focused on two specific aspects of
this geometry: the expected dimension of the phase structure in (usually
6-dimensional) phase space, and the classification of singularities when 
a $D$-dimensional phase structure is projected into a $D$-dimensional survey.

Phase mixing of disrupted small, hot galaxies generally leads to 3-dimensional
phase structures but the dimensionality can be smaller if the potential of the
host galaxy has special symmetries (e.g. spherical or Keplerian), or if the
eigenvalues of the Hessian (eq. \ref{eq:hhh}) are very different in
magnitude. Phase mixing of large, cold galaxies leads to 2-dimensional phase
structures. 

The structurally stable singularities in $D$-dimensional surveys are folds
($D\ge 1$), cusps ($D\ge 2$), swallowtails, elliptic umbilics and hyperbolic
umbilics ($D\ge 3$). 

Even though phase mixing is a simple process, there remain many unresolved
theoretical issues. How fast do small-scale gravitational irregularities or
large-scale orbital chaos disrupt phase structures? Do phase structures
provide a significant source of relaxation in galaxies (\cite{TO98})? What are
the properties of phase structures in galaxies that we expect from standard
models of structure formation? What statistical measures can we use to
characterize phase structures and their projections?


\begin{thebibliography}{}

\bibitem[Arnold 1986]{A86} Arnold, V. I. 1986, Catastrophe Theory, 2nd
ed. (Berlin: Springer)

\bibitem[Arnold and Avez 1968]{AA68} Arnold, V. I., and Avez, A. 1968, Ergodic
Problems of Classical Mechanics (New York: W. A. Benjamin)

\bibitem[Berry and Upstill 1980]{BU80} Berry, M. V., and Upstill, C. 1980,
Progress in Optics 18, ed. E. Wolf (Amsterdam: North-Holland), 257

\bibitem[Binney and Tremaine 1987]{BT87} Binney, J., and Tremaine, S. 1987,
Galactic Dynamics (Princeton: Princeton University Press)

\bibitem[Dehnen 1998]{D98} Dehnen, W. 1998, AJ 115, 2384

\bibitem[Eggen 1965]{E65} Eggen, O. J. 1965, in Galactic Structure, Stars and
Stellar Systems 5, eds. A. Blaauw and M. Schmidt (Chicago: University of
Chicago Press), 111

\bibitem[Gilmore 1981]{G81} Gilmore, R. 1981, Catastrophe Theory for Scientists
and Engineers (New York: Wiley)

\bibitem[Hernquist and Quinn 1988]{HQ88} Hernquist, L., and Quinn, P. J. 1988,
ApJ 331, 682

\bibitem[Ibata et al. 1997]{IWG97} Ibata, R. A., Wyse, R.F.G., Gilmore, G.,
Irwin, M. J., and Suntzeff, N. B. 1997, AJ, 113, 634

\bibitem[Irwin and Hatzidimitriou 1995]{IH95} Irwin, M., and Hatzidimitriou,
D. 1995, MNRAS 277, 1354

\bibitem[Johnston et al. 1996]{JHB96} Johnston, K. V., Hernquist, L., and
Bolte, M. 1996, ApJ 465, 278

\bibitem[Lynden-Bell and Lynden-Bell 1995]{LB95} Lynden-Bell, D., and
Lynden-Bell, R. M. 1995, MNRAS 275, 429

\bibitem[Majewski et al. 1996]{MHM96} Majewski, S. R., Hawley, S. L., and
Munn, J. A. 1996, in Formation of the Galactic Halo$\ldots$Inside and Out,
ed. H. Morrison and A. Sarajedini (San Francisco: ASP), 119

\bibitem[Poston and Stewart 1978]{PS78} Poston, T., and Stewart, I. N. 1978,
Catastrophe theory and its applications (London: Pitman)

\bibitem[Sikivie and Ipser 1992]{SI92} Sikivie, P., and Ipser J. 1992,
Phys. Lett. B, 291, 288
 
\bibitem[Tremaine and Ostriker 1998]{TO98} Tremaine, S., and Ostriker,
J. P. Submitted to MNRAS

\bibitem[Zeeman 1977]{Z77} Zeeman, E. C. 1977, Catastrophe Theory: selected
papers 1972--1977 (Reading: Addison-Wesley) 
\end{thebibliography}
\end{document}